\documentclass[review,12pt]{elsarticle}

\usepackage{lineno,hyperref}
\modulolinenumbers[5]

\journal{Annals of Physics}

\usepackage{amsmath}
\DeclareMathOperator{\tr}{Tr}

\usepackage{color}
\usepackage[dvipsnames]{xcolor}
\usepackage{graphicx}
\usepackage{caption}
\usepackage{subcaption}
\begin{document}

\begin{frontmatter}

\title{Phase collapse and revival of a 1-mode Bose-Einstein condensate induced by an off-resonant optical probe field and superselection rules}



\author{L. G. E. Arruda }
\address{Departamento de F\'{\i}sica, Universidade Federal de S\~{a}o Carlos, 13565-905, S\~{a}o Carlos, SP, Brazil}

\author{G. A. Prataviera}
\address{Departamento de Administra\c{c}{\~a}o, FEA-RP, Universidade de S{\~a}o Paulo, 14040-905, Ribeir\~{a}o Preto, SP, Brazil}

\author{M. C. de Oliveira}
\address{Instituto de F\'{\i}sica Gleb Wataghin, Universidade de Campinas, 13083-970, Campinas, SP, Brazil}

\cortext[mycorrespondingauthor]{Corresponding author}
\ead{marcos@ifi.unicamp.br}

\begin{abstract}

Phase collapse and revival for Bose-Einstein condensates is a nonlinear phenomena appearing due atomic collisions. While it has been observed in a general setting involving many modes, for one-mode condensates its occurrence is forbidden by the particle number superselection rule (SSR), which arises because there is no phase reference available. We consider a single mode atomic Bose-Einstein condensate interacting with an off-resonant optical probe field. We show that the condensate phase revival time is dependent on the atom-light interaction, allowing optical control on the atomic collapse and revival dynamics. Incoherent effects over the condensate phase are included by considering a continuous photo-detection over the probe field. We consider conditioned and unconditioned photo-counting events and verify that no extra control upon the condensate is achieved by the probe photo-detection, while further inference of the atomic system statistics is allowed leading to a useful test of the SSR on particle number and its imposition on the kind of physical condensate state.
\end{abstract}


\end{frontmatter}



\section{Introduction}

Collapse and revival phenomena in Bose-Einstein condensates (BECs) have been investigated since 1996 \cite{You05}, shortly after its experimental achievement with ultracold atoms \cite{Anderson, Davis01}. They are a consequence of the quantized structure of the matter field and the coherent interactions between the atoms through atomic collisions (See \cite{Greiner}, \cite{bach}, and references therein). 

Dynamically originated due the presence of nonlinearities, phase collapse and revival is similar in nature to the collapse and revival appearing in Rabi oscillations when a single two-level atom interacts with a single mode of a quantized optical field in the so called Jaynes-Cummings model \cite{jaynes}, or when a coherent light field propagates in a non-linear medium \cite{Scully1997}. The Jaynes-Cummings model was one of the first totally quantized and exactly solvable models of interaction between matter and radiation showing non-trivial features. The collapse and revival effects in the population difference between the states of a two-level atom placed in a single mode optical cavity was one of the first predicted and still explored phenomena \cite{shore, berman}, revealing quantum properties of a radiation field. It was observed experimentally for the first time using a microwave mode \cite{rempe02}. Achievement of Bose-Einstein condensation led to a new platform to explore such coherent phenomena considering the interaction of  quantized optical fields with ultracold many particle atomic systems \cite{Ritsch}.

The possibility of coexistence of nonlinear effects due to external influences opens an interesting avenue for investigating and controlling \cite{John} such phenomena in BECs. Particularly, the control of phase collapse in BECs is extremely relevant for atomic interferometry since it might prevent phase diffusion, while allowing a performance below the standard quantum limit \cite{Gross2010}. Several attempts on both developing new tools and pushing forward the possible frontiers for collapse control have been investigated for double-well condensates (See {\cite{milburn03}, \cite{milburnbook}}, \cite{John} and references therein). A possible approach is to consider the action of an external light probe over the atomic system, although, in general this leads to an even worse situation where the light field itself induces phase collapse in an irreversible manner. It is expected that, unless a time dependent interaction is employed, no further control is achieved.

Nonetheless, the interaction of the atomic system with a quantized light probe field can be engineered with the assistance of an additional pump field (See \cite{Ritsch} for an excellent review on this topic). This scheme allows, e.g., the simultaneous amplification of atomic and optical fields, as well as control of the atomic field statistical properties \cite{Moore01,Moore02, Vivian}.
Previously, with a setup relying on the atomic system-probe field interaction mediated through a classical pump \cite{nosso}, we showed that under continuous photo-counting, the moments of the probe light photon number might carry information about the even moments of the atom number. However, neither the possibility of controlling of atomic properties with the proposed setup, or a detailed analysis about the effects of conditioned single photo-counting events over the BEC state, was carried out. Such analysis is important, given the interest in phase collapse time, since in some situations the detection process allows an additional control or better inference of parameters. For example, in \cite{saba} an optical probe continuous detection allows one to create relative phase of two spatially separated atomic BEC. Also, as showed in \cite{Diehl}, even dissipative environments can be useful to tailor dynamics of states and phase in cold atomic samples \cite{dalton3}.

{We should remark that the phenomenon of collapse and revival for a single mode condensate overlaps with another significant issue in many-particle physics, which is the particle number Super-Selection Rule (SSR)\cite{wick}. The imposition of a global particle number conservation forbids the one-mode (many-particle) state to exist in a superposition of Fock states - only single Fock states or a mixture of those are allowed. However neither a Fock state nor a mixture of Fock states have a well defined phase, and therefore there will be no phase collapse and revival dynamics. A reference frame is needed to confirm that a quantum state which violates superselection rule exists \cite{Susskind}, (such as a Glauber coherent state in the context of the particle number SSR). Therefore, although this might be less restrictive when dealing with two or many-modes condensates, see for example the discussion in \cite{bach}, due to the possibility of creating states with a well-defined relative phase in two mode systems \cite{dalton3} for example, for one-mode condensates this is a severe constraint. Both for systems of identical massive particles (such as bosonic atoms) and for systems of identical massless particles (such as photons) there is however a long standing discussion of whether the SSR  on particle number actually applies \cite{leggett,dalton1,dalton2,bartlett1,bartlett,molmer,sanders}. For systems of identical massive particles there are strong reasons why it does (see \cite{leggett,dalton1,bartlett1}), whereas for systems of identical massless particles the situation is less clear (see \cite{dalton1,molmer,sanders}). In Ref \cite{dalton1} (see Sections 3.2, 3.3, 3.4 and Appendices K, L ) the reasons why the situation may differ for atoms in Bose condensates and photons in optical modes are discussed. We see the possibility of observing the collapse and revival of the one-mode condensate as a test for the particle number SSR.}

To address these questions, in this paper we analyse the situation in which a single mode atomic BEC is coupled to a quantized optical probe field, through an undepleted optical pump. The nonlinear nature of the engineered interaction between the atoms and the probe field, induces a proper dynamics of collapse and revival of the atomic phase, even in the absence of atomic collisions for a very diluted atomic gas, {if a state with a well defined initial phase is assumed}. We show that the characteristic revival time, depends on the commensurability between the parameters such as the coupling constants and the detuning between the transition frequency of the two level atoms (that represents our BEC) and the frequency of the pump field, as well as, the atomic collision parameter. This allows control over the atomic collapse and revival through the effective atom-light coupling parameters. However for states with not well defined initial phases no collapse and revival control is allowed. In addition, we include continuous photo-counting over the probe field and analyse its possible control over the collapse and revival times. We conclude that no further control is achieved by these measurements {even when a state with a well defined phase is assumed. However since the optical field collects information about the BEC collapse and revival dynamics, when the photo-counting statistics is contrasted with the optical field phase dynamics, one gets an unequivocal proof of the condensate mode collapse and revival together with an inference on the BEC state}.

{ This discussion is relevant for experimental investigation on the super-selection rules and particle number conservation. So in similar fashion to many modes \cite{bach}, although coherent states would be inadequate to describe a single mode condensate with total number of atoms fixed (if no additional system is introduced to define a phase reference), it reveals many relevant features of the system dynamics. Indeed if collapse and revival is to be observed at all for a single BEC mode, the only possibility is that it must exist in a superposition of Fock states, violating the SSR.}

The paper is organised as follows: In Section 2 we deduce the model for a BEC trapped in a single well potential, interacting with a classical undepleted optical field and a quantized optical probe mode in the far-off resonant regime. In Section 3 we discuss on the possible initial system states satisfying or violating the SSR. In Section 4, we analyse the collapse and revival dynamics of the BEC phase using the Husimi function and the variance of the phase operator. We show how the coupling parameters between atoms and light determine and enable to control the collapse and revival dynamics. In Section 5 we include an incoherent process through a continuous photo-detection on the quantized optical probe field. The effect of optical photo-counting over the BEC phase dynamics is analysed {and a discussion on the SSR on the condensate mode and its statistics is given}. Finally, in Section 6 we present the conclusions.

\section{Model}

We consider a system of bosonic two-level atoms interacting via two-body collisions and coupled through electric-dipole interaction with two single-mode running wave optical fields of frequencies $\omega _{1}$ and $\omega _{2}$, and wavevectors $\mathbf{k}_{1}$ and $\mathbf{k}_{2}$, respectively. The probe optical field (1) is treated quantum mechanically, while the pump (2) is undepleted and is treated classically. Both fields are assumed to be far off-resonance from any electronic transition, and the excited state population is small so that spontaneous emission may be neglected. Similarly, collisions among excited state atoms, as well as, collisions between ground state atoms with excited state ones, are very improbable and can also be neglected. In this regime the excited state can be adiabatically eliminated and the ground state atomic field plus the optical probe evolve coherently under the effective Hamiltonian \cite{Moore02,Vivian,nosso} 
%
\begin{eqnarray} 
	{H} &=&\int d^{3}\mathbf{r} {\Psi}^{\dagger }(\mathbf{r})\left[
	{H}_{0}+\frac{U}{2}{\Psi}^{\dagger }(\mathbf{r}) {\Psi}(%
	\mathbf{r})+ \hbar\frac{\left|g_{2}\alpha_{2}\right|^2}{\Delta}\right] {\Psi}(
	\mathbf{r})\nonumber \\
	&&+\hbar \int d^{3}\mathbf{r} {\Psi}^{\dagger }(\mathbf{r})\left[ \frac{g_{1}^{\ast }g_{2}\alpha_{2}}{\Delta } {a}%
	_{1}^{\dagger }e^{-i\mathbf{k\cdot r}}+ \text {H.c.}
	\right]  {\Psi}(%
	\mathbf{r}) \nonumber \\
	&&+\hbar \left[\left(\omega_{1} - \omega_{2}\right) +\frac{\left|g_{1}\right|^2}{\Delta}\int d^{3}\mathbf{r\;}
	{\Psi}^{\dagger }(\mathbf{r}) {\Psi}(\mathbf{r})\right] {a}_{1}^{\dagger } {a}_{1}.
	\label{eq1}
\end{eqnarray}
%
${\Psi}\left(\mathbf{r}\right)$ is the ground state atomic field operator which satisfies the usual bosonic commutation relations $\left[{\Psi}\left(\mathbf{r}\right),{\Psi}^{\dag}\left(\mathbf{r}\ '\right)\right] = \delta\left(\mathbf{r} - \mathbf{r}\ '\right)$. $ {H}_{0}=-\frac{\hbar ^{2}}{2m}\nabla ^{2}+V(\mathbf{r})$ is the trapped atoms Hamiltonian, where $m$ is the atomic mass and $V(\mathbf{r})$ is the trap potential. $\Delta = \omega_2 -\nu$ is the detuning between the atomic transition and the optical pump frequencies, $g_{1}$ and $g_{2}$ are the atom-light coupling coefficients, and $\mathbf{k}=\mathbf{k}_{1}-\mathbf{k}_{2}$.  The operators $ {a}_{1}$ and $ {a}^{\dag}_{1}$ (given in a rotating frame with frequency $\omega _{2}$) satisfy the commutation relation $\left[ {a}_{1},  {a}_{1}^{\dag}\right] = 1$, and  $\alpha_{2}$ is the pump amplitude. Two-body collisions were included in the s-wave scattering limit, where  $U=\frac{4\pi \hbar ^{2}a}{m}$ and $a$ is the s-wave scattering length \cite{Pethick}. 

We expand the atomic field operators in terms of the orthogonal set of trap eigenmodes $\left\{ \varphi _{n}(\mathbf{r})\right\}$, as
\begin{equation}
	{\Psi}(\mathbf{r})=\sum_{n} {c}_{n}\, \varphi_{n}(\mathbf{r}), \label{Expansion_of_atomic_field_operators}
\end{equation} 
where the trap eigenmodes satisfies the orthogonality relation $\int d^{3}\mathbf{r\;}\varphi _{m}^{\ast }(\mathbf{r})\varphi _{n}(\mathbf{r})=\delta _{mn}$ and the eigenvalue equation ${H}_{0}\varphi _{n}(\mathbf{r})=\hbar \widetilde{\omega}_{n}\varphi _{n}(\mathbf{r})$, where $\widetilde{\omega}_{n}$ are the corresponding trap eigenfrequencies. ${c}_{n}$ is the atom annihilation operator in the mode $n$, and together with the creation operator, satisfies the regular commutation relation $\left[ {c}_{n}, {c}_{n}^{\dagger}\right]=\delta_{nm}$. 

With the expansion (\ref{Expansion_of_atomic_field_operators}), the Hamiltonian (\ref{eq1}) takes the following form
\begin{eqnarray}
	{H} &=& \hbar\sum_{n}\left(\tilde{\omega}_{n}+\frac{\left|g_{2}\alpha_{2}\right|^2 }{\Delta}\right){c}_{n}^{\dagger }{c}_{n}
	+\hbar\sum_{ijlm}\kappa_{ijlm} {c}_{i}^{\dagger }{c}_{j}^{\dagger }{c}_{l}{c}_{m} \nonumber \\
	&+&\hbar\sum_{m\,n}\left(\frac{g_{1}^{\ast }g_{2}\alpha_{2}}{\Delta}\chi_{mn}{a}_{1}^{\dagger }{c}_{n}^{\dagger }{c}_{m}+\frac{g_{2}^{\ast }g_{1}\alpha_{2}^{\ast }}{\Delta
	}\chi_{nm} a_{1}{c}_{m}^{\dagger }{c}_{n} \right) \nonumber \\
	&+& \hbar \left(\delta +\frac{\left|g_{1}\right|^2}{\Delta}\sum_{n}{c}_{n}^{\dagger }{c}_{n}\right){a}_{1}^{\dagger }{a}_{1}, \label{total}
\end{eqnarray}
where
\begin{equation}
	\kappa_{ijlm}=\frac{U}{2\hbar}\int d^{3}\mathbf{r\;}\varphi _{i}^{\ast }(\mathbf{r
	})\varphi _{j}^{\ast }(\mathbf{r})\varphi _{l}(\mathbf{r})\varphi _{m}(\mathbf{r}),
\end{equation}
is the inter-modes collision parameter and
\begin{equation}
	\chi_{nm}=\int d^{3}\mathbf{r\;}\varphi _{m}^{\ast }(\mathbf{r})e^{-i\mathbf{\mathbf{k}\cdot r}}\varphi _{n}(\mathbf{r}),
\end{equation}
is the optical transition matrix elements from the  $m$th state to $n$th state, and $\delta=\omega_{1} - \omega_{2}$. 

Hamiltonian (\ref{total}), accounts for the process of atomic and optical parametric amplification due momentum exchange between atoms and optical field \cite{Moore02}. Here, we assume a sufficiently low temperature so that all atoms form a pure BEC in the trap ground state, and to avoid 
scattering process which transfers atoms to other trap modes via momentum exchange we consider that 
$\mathbf{k}_{1}\approx \mathbf{k}_{2}$, and for simplicity that $\omega_{1} \approx \omega_{2}$\footnote{Otherwise, its only effect is an energy shift in the probe field that does not depend on the atom number and do not change at all our results. Besides, the two optical modes can yet be distinguished by their polarizations.}. With these assumptions, only the atomic operators $c^{\dag}_{0}$ ($c_{0})$ corresponding to creation (annihilation) of atoms at the trap ground state remains, and the Hamiltonian (\ref{total}) simplifies to
\begin{eqnarray}
	{H} &=& \hbar\left(\widetilde{\omega}_{0}+\frac{\left|\widetilde{g}_{2}\right|^2 }{\Delta}\right) {c}_{0}^{\dagger } {c}_{0}
	+\hbar{\kappa}{c}_{0}^{\dagger } {c}_{0}^{\dagger } {c}_{0} {c}_{0} \nonumber \\
	&&+\hbar  {c}_{0}^{\dagger } {c}_{0}\left(\frac{g_{1}\widetilde{g}_{2}^{\ast }}{\Delta
	}  {a}+\frac{\widetilde{g}_{2}g_{1}^{\ast }}{\Delta} {a}^{\dagger } \right)
	+ \hbar \frac{\left|g_{1}\right|^2}{\Delta} {c}_{0}^{\dagger } {c}_{0} {a}^{\dagger } {a},\label{hamiltonian}
\end{eqnarray}
where $\widetilde{g}_{2}=g_{2}\alpha_{2}$ and $\kappa=\frac{U}{2\hbar}\int d^{3}\mathbf{r\;}|\varphi _{0}(\mathbf{r})|^4$ is the collision parameter between the atoms in the trap ground state. The optical probe mode index was dropped in order to simplify the notation. 

{
\section{Particle number superselection rule and allowed joint system state}

The dynamical behaviour of the joint light-atom system and presence or not of the BEC phase collapse and revival is strongly dependent on the initial state. The local conservation of atom number for the BEC imposes a restriction on the kind of possible physical states. It has been a long standing question \cite{wick} that additional rules must be imposed when dealing with systems of indistinguishable particles - the so-called super selection rules (SSR). SSR and particle conservation number are intimately related to the requirement of a reference frame \cite{bartlett,dalton1,dalton2} for definition of state with well defined phase, forbidding an isolated one-mode multiparticle system to exist in a coherent superposition of Fock states. Therefore in a light-atom system, or the joint state is entangled - which allows to the matter field state to have a well defined phase (relative to the optical field), or if separable the BEC state must be a Fock state or a mixture of those. This issue has been discussed for a long time, but is not totally settled, since theoretically it might be possible to devise a preparation scheme for preparing a BEC in a single mode with a well-defined phase relative to another system which acts as a phase reference, though whether the high frequency oscillations of the coherences between different particle number states would be observable in non-relativistic quantum physics is doubtful (see \cite{dalton1}). We here take a pragmatic investigative point of view, by assuming both situations, in agreement with the superselection rule or violating it, to check the consistency of information about the condensate that can be acquired by detection of the optical probe field.
We assume that initially the atoms plus optical probe field joint state is completely disentangled $\rho\left(0\right) = \rho_{A}\left(0\right)\otimes\rho_{L}\left(0\right)$, where $A$ and $L$ stand for atoms and light, respectively. {Since the optical field interacts with the atomic system inside a cavity, the initial optical field  can be prepared in a coherent Glauber state given by $\rho_{L}\left(0\right) = \vert\beta\rangle\langle\beta\vert$.} The time evolution of the combined system can be obtained exactly from the Hamiltonian (\ref{hamiltonian}). In the next section we will present an analysis of the BEC and light phase dynamics considering both pure and mixed initial states for the atomic field. In particular, for a hypothetical  pure atomic initial state given by a coherent superposition of number operator eigenvalues, $\vert\psi\left(0\right)\rangle_{A}=\sum_{m}C_{m}\vert m\rangle$, we address a way to observe and control the revival times of the BEC phase by tuning the effective coupling between atoms and the probe field. In that situation the evolution of the initial state $\vert\psi\left(0\right)\rangle = \sum_{m}C_{m}\vert m\rangle\otimes\vert\beta\rangle$, is given by (See the appendix for a detailed derivation) 
\begin{equation}
	\vert\psi\left(t\right)\rangle = \sum_{m}C_{m}e^{\Phi_{m}\left(t\right)}\vert m\rangle\otimes\vert\beta_{m}\left(t\right)\rangle, \label{evaluated_state}
\end{equation}
where
\begin{equation}
	\beta_{m}\left(t\right) = \beta e^{-i\frac{\left|g_{1}\right|^{2} m}{\Delta}t} + \frac{\widetilde{g}_{2}}{g_{1}}(e^{-i\frac{\left|g_{1}\right|^{2} m}{\Delta}t} -1), \label{beta}
\end{equation} 
is the probe field coherent state time dependent amplitude and
\begin{eqnarray}
	\Phi_{m}\left(t\right) &=&\frac{1}{2}\left[\left|\beta_{m}\left(t\right)\right|^{2} - \left|\beta\right|^{2}\right] + \frac{\widetilde{g}_{2}^*}{g_{1}^*}\left[\beta_{m}\left(t\right) -\beta\right] \nonumber \\
	&-& i\left(\widetilde{\omega}_{0}+\frac{|\widetilde{g}_{2}|^2}{\Delta}\right)mt - i\kappa m\left(m-1\right)t, \label{phi}
\end{eqnarray}
is the relative phase introduced by the dynamics. The density operator for the combined system is $ {\rho}\left(t\right)=\vert\psi\left(t\right)\rangle\langle\psi\left(t\right)\vert$ from which we can obtain both: the BEC reduced density operator ${\rho}_{A}\left(t\right)= \tr_{L} {\rho}\left(t\right)$ or the optical field reduced density operator ${\rho}_{L}\left(t\right)= \tr_{A} {\rho}\left(t\right)$. 

It also follows that for an initial arbitrary mixture $\rho_{A}(0)=\sum_{m,n}\rho_{A_{n,m}}\vert m\rangle\langle n\vert$ the total state is given by the density operator
\begin{equation}
	{\rho}\left(t\right) = \sum_{m,n}\rho_{A_{n,m}}e^{\Phi_{m}\left(t\right)+\Phi^{*}_{n}\left(t\right)}\left\vert m\right\rangle\left\langle n\right\vert \otimes\left\vert \beta_{m}\left(t\right)\right\rangle\left\langle \beta_{n}\left(t\right)\right\vert,
	\label{evaluated_state2}
\end{equation}
which for a incoherent statistical mixture writes as \begin{equation}
\rho_{A_{n,m}}=P_{m}\delta_{m,n},\end{equation} and only the diagonal terms in Eq. (\ref{evaluated_state2}) remain.}

\section{Condensate phase dynamics}

{As emphasised in the last section, to analyse the condensate phase dynamics and propose a scheme to control the phase revival times through the model described above, we examine three different types of atomic initial states. First we look to the case where the initial single mode BEC state is described by a coherent superposition of number operator eigenvalues given by the coherent Glauber state.
Then we consider a well defined number state to be the initial single mode BEC state. Also, we consider a statistical mixture of states with differing particle numbers with a binomial weight, which is a reasonable state to consider as a simple example. 

Considering that the initial state of the atoms is a coherent Glauber state we have $C_{m} = e^{-\frac{\vert\alpha\vert^{2}}{2}}\frac{\alpha^{m}}{\sqrt{m!}}$ in Eq. (\ref{evaluated_state}). In this particular case,} analyzing each term from Hamiltonian (\ref{hamiltonian}) separately, we are able to visualise different scenarios. For instance, its second term plays a key role on the collapse and revival of the atomic phase dynamics, typical of one-mode BECs (See Fig. 1 in Greiner et al. \cite{Greiner} for the atomic Husimi function $Q_{A}\left(\alpha_{A},t\right)=\frac{1}{\pi}\langle\alpha_{A}\vert {\rho}_{A}\left(t\right)\vert\alpha_{A}\rangle$ corresponding here to $\frac{\vert g_{1}\vert^{2}}{\Delta}\ll\kappa$). After evolving into states of totally uncertain phases and also some exact superpositions of coherent states, the atomic state fully recover the initial phase at the revival time $t_{rev}^{C}= \pi/\kappa$.

\begin{figure}[ht]\par\includegraphics[scale=0.22]{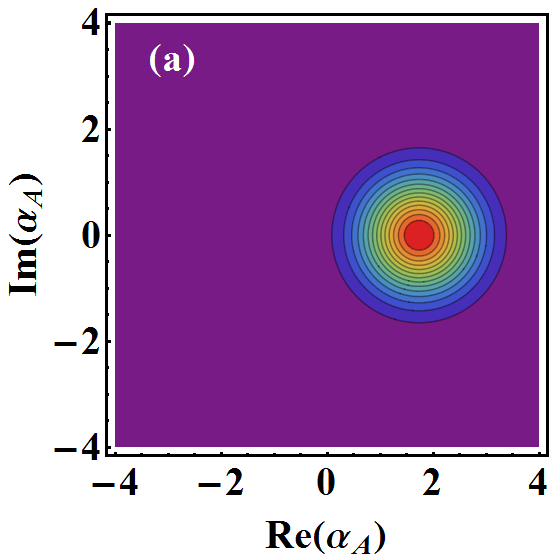}\includegraphics[scale=0.22]{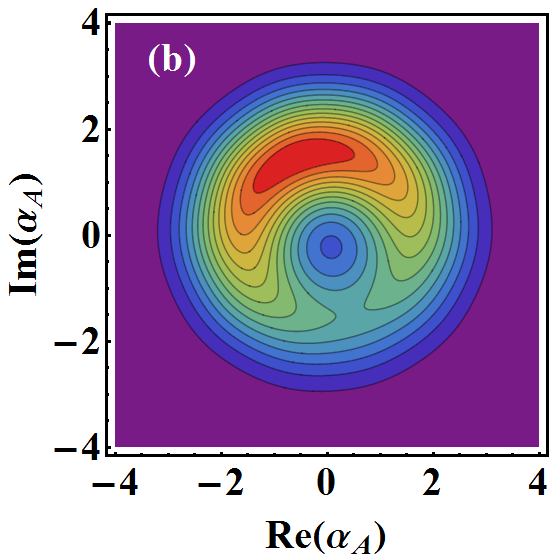}\includegraphics[scale=0.22]{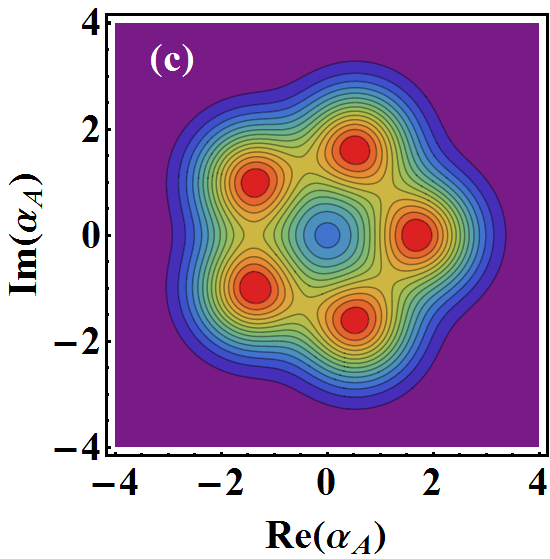}\par\includegraphics[scale=0.22]{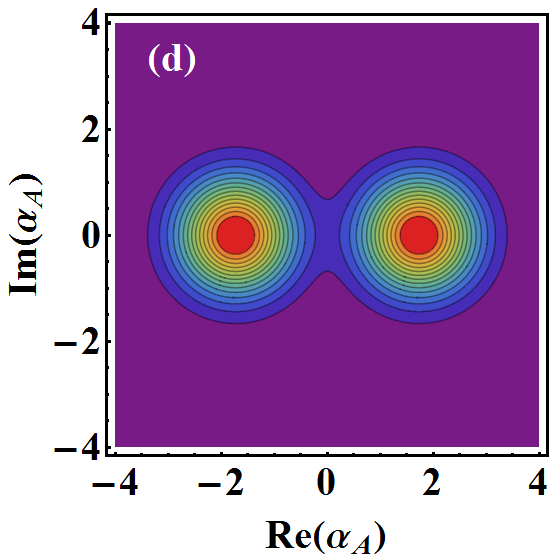}\includegraphics[scale=0.22]{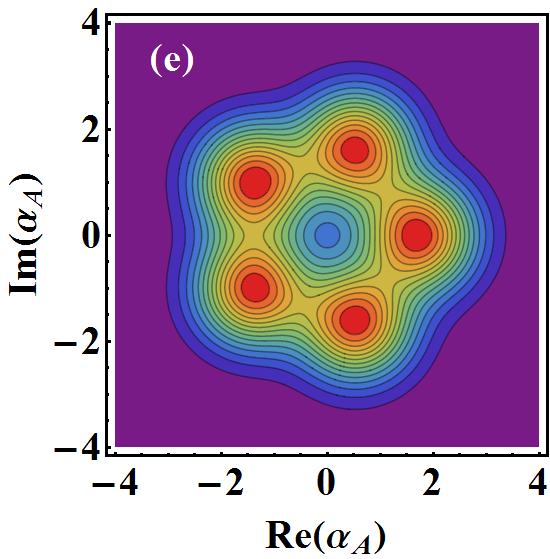}\includegraphics[scale=0.22]{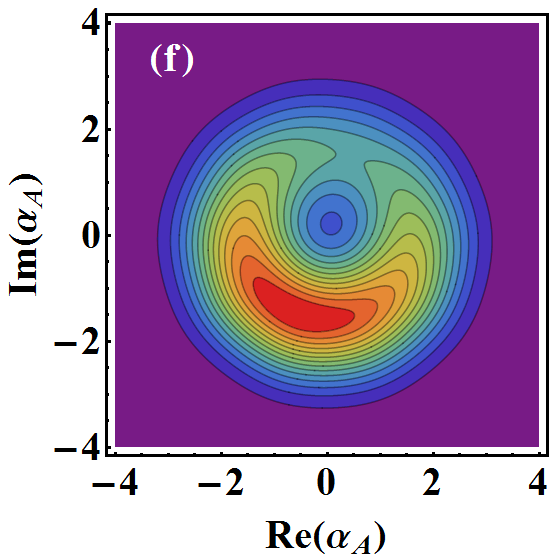}
	\caption{Husimi function dynamics of the atomic density operator for $\vert\alpha\vert^{2}=3$ in the limit $\frac{\vert g_{1}\vert^{2}}{\Delta}\gg\kappa$. In \textbf{(a)}  $t=0$; \textbf{(b)} $t=0.1(2\pi\Delta/{\vert g_{1}\vert^{2}})$; \textbf{(c)} $t=0.4(2\pi\Delta/{\vert g_{1}\vert^{2}})$; \textbf{(d)} $t=0.5(2\pi\Delta/{\vert g_{1}\vert^{2}})$; \textbf{(e)} $t=0.6(2\pi\Delta/{\vert g_{1}\vert^{2}})$; \textbf{(f)} $t=0.9(2\pi\Delta/{\vert g_{1}\vert^{2}})$. A full revival of the initial coherent state in  \textbf{(a)} is reached at $t^{L}_{rev}= 2\pi\Delta/{\vert g_{1}\vert^{2}}$. We set the probe field initial intensity as $\vert\beta\vert^{2}=3$.}\label{Image_1}
\end{figure}

The third and fourth terms in Hamiltonian (\ref{hamiltonian}) describe the couplings between atoms and the quantum optical probe. The third term corresponds to a transfer of photons from the classical undepleted pump optical field to the quantized probe mode mediated by the atoms. The fourth term is quite relevant since it will also contribute to the collapse and revival of the BEC phase (and of the optical probe) due to the cross-Kerr type of nonlinearity. The regime $\frac{\vert g_{1}\vert^{2}}{\Delta}\gg\kappa$, for a small detuning (large enough though to prevent spontaneous emission \cite{Ritsch}) or for a very diluted atomic gas ($\kappa\approx 0$), is depicted in Fig. \ref{Image_1} for $\vert\alpha\vert^{2}=3$. The collapse and revival dynamics occurs at a completely distinct revival time $t_{rev}^{L}={2\pi\Delta}/{|g_{1}|^{2}}$, depending on the atom-light interaction parameter.

\begin{figure}[!ht]\begin{center}\includegraphics[scale=.5]{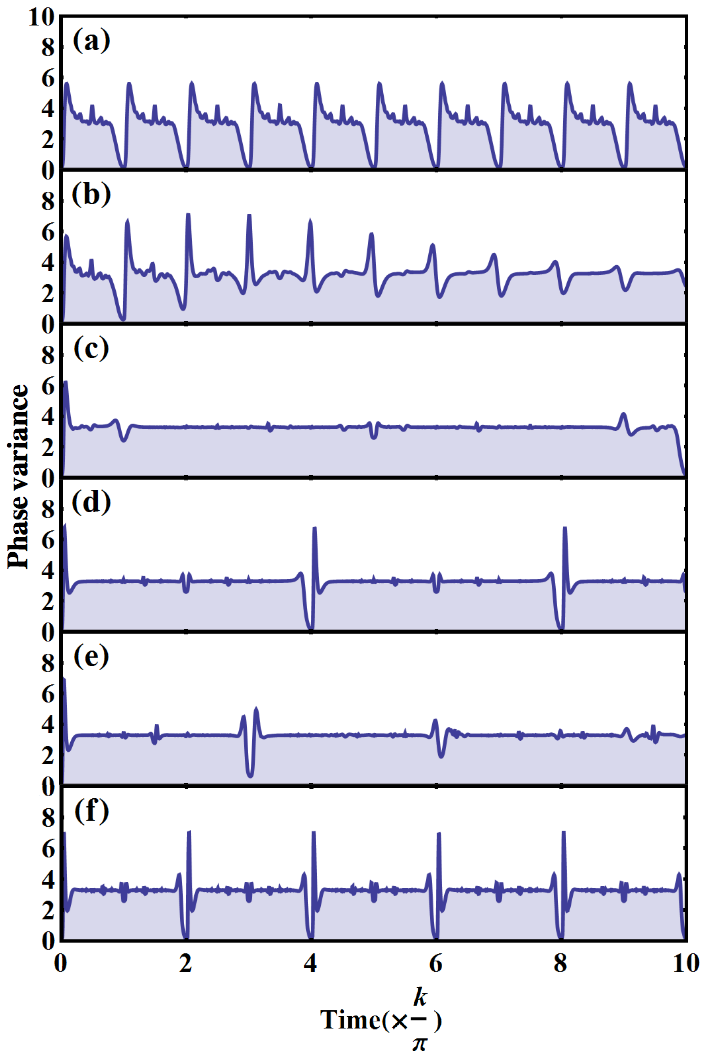}\end{center}\caption{Phase variance for the BEC owing to both atom collisions and interaction with the optical probe. The latter one disturbs enormously the revival time, when the atomic state returns approximately to a coherent state.  From top to bottom, (a) $|g_{1}|^2/\kappa\Delta=0$, (b) $|g_{1}|^2/\kappa\Delta=1/50$, (c) $|g_{1}|^2/\kappa\Delta=1/5$, (d) $|g_{1}|^2/\kappa\Delta=1/2$, (e) $|g_{1}|^2/\kappa\Delta=2/\pi$, and (f) $|g_{1}|^2/\kappa\Delta=1$. We set the mean number of atoms and the probe field initial intensity as  $\vert\alpha\vert^{2}=3$ and  $\vert\beta\vert^{2}=3$, respectively.} \label{phase}
\end{figure}

A nontrivial dynamics occurs when both the second and fourth therms in (\ref{hamiltonian}) are of the same order, which can be reached by varying the detuning $\Delta$. In a general way the revival occurs whenever the terms in a expansion spanned by Fock states of the atomic state are in phase. While it is easy to describe this when only the collision term or the interaction with the optical probe are on, the situation is more complicated when both terms are relevant. To show that, let us analyse the behaviour of the variance of the phase operator for the atomic mode \cite{milburnbook}. For an initial coherent state of a large amplitude $\alpha$, the phase variance is approximately $V(\phi)=1/4|\alpha|^{2}$ and clearly shows that the phase is well defined for large $\alpha$. For the chosen amplitude of $\vert\alpha\vert^{2}=3$ the phase revival, as depicted in Fig. \ref{phase}, occurs every time $V(\phi)$ approaches zero. Note that the collapse and revival dynamics similar to the one in Fig. 1 in Greiner et al. \cite{Greiner} repeats several times in Fig. \ref{phase}(a) for the atomic mode at the chosen time scale. However this behavior is very fragile and changes completely with the inclusion of a very small perturbation in $\left\vert{g_{1}}\right\vert^2/\kappa \Delta$. The revival time for this scenario can be investigated by expanding the collision and interaction terms in a Fock basis, $\{|n\rangle,|m\rangle\}$ for the atoms and light field, respectively. The revival time is  not dependent on the third term of the Hamiltonian (\ref{hamiltonian}), whose effect is only to displace an initial light field coherent state depending on the number of atoms in the BEC. Then, considering a rotating frame with frequency $\widetilde{\omega}_{0}+\frac{\left|\widetilde{g}_{2}\right|^2 }{\Delta}$, the revival time coincides with the recurrence of the initial phase, which will take place whenever the following relation is satisfied
\begin{equation}
	e^{-i\left[\kappa n\left(n-1\right) + \frac{\vert g_{1}\vert^{2}}{\Delta}nm\right]t} = e^{-2il\pi},    
\end{equation}
where $l$,  $m$, and  $n$ are positive integers. This occurs for times such that
\begin{equation}
	t_{rev}=\frac{2\kappa\Delta}{|g_1|^ 2}\left[\frac{l}{\frac{\kappa\Delta}{|g_1|^2}n(n-1)+nm}\right]t_{rev}^C,\label{trev}
\end{equation}
since $t_{rev}^{C}= \pi/\kappa$. The revival depends on the commensurability between $|g_1|^2/\Delta$ and $\kappa$. Since $n$ and $m$ are integers, when $|g_1|^2/\kappa\Delta$ is rational ($\equiv\frac{p}{q}$, with $p$, $q$ integers) we have
\begin{equation}
	t_{rev}=\frac{2\kappa\Delta}{|g_1|^ 2}\left[\frac{lq}{pn(n-1)+qnm}\right]t_{rev}^C.\label{trev2}
\end{equation}
Given that ${pn(n-1)+qnm}$ is an integer and $l$ is arbitrary, there always exists a $lq={pn(n-1)+qnm}$ in the numerator of Eq. (\ref{trev2}), so that the revival time reduces to $t_{rev}=2\frac{\kappa\Delta}{|g_1|^ 2}t_{rev}^C$. However, for an irrational $|g_1|^2/\kappa\Delta$, there is no revival at all.

Let us exemplify with the inclusion of perturbations in the interaction with the light field (by decreasing the detuning). In Fig. \ref{phase}(b), for $\left\vert{g_{1}}\right\vert^2/\kappa \Delta=1/50$, we see an inhibition of the number of revivals at this time scale, even though an actual revival occurs at a different time scale at $100\; t_{rev}^C$. 
In a similar way Fig. \ref{phase}(c) for $|g_1|^2/\kappa\Delta=1/5$, shows that the revival time is  $10\; t_{rev}^C$, and in
Figs. \ref{phase}(d) and \ref{phase}(f), for $|g_1|^2/\kappa\Delta=1/2$, and $|g_1|^2/\kappa\Delta=1$, where the revival time is $4\; t_{rev}^C$ and $2\; t_{rev}^C$, respectively. However, for $\frac{g^2}{\kappa\Delta }=\frac{2}{\pi}$, as in Fig. \ref{phase}(e), there is no phase revival, as we expected.
The relevant aspect on  the atomic revival time change is the possibility to control it through the variation of $\Delta$. Whenever the revival occurs, and only then, both optical pump and the BEC are left disentangled. Besides the control over the revival, one could be interested in this available entanglement. Several entangled coherent states occur, being a typical example at exactly half of the revival time -  the optical probe and BEC are approximately left in a state
\begin{equation}|\psi\rangle\approx|\beta\rangle|\alpha_{+}\rangle+|\beta\rangle|\alpha_{-}\rangle,
\end{equation}
where $|\alpha_{\pm}\rangle$ are odd and even coherent states. Such states are useful, \textit{e.g.}, for teleportation \cite{tele}.

\begin{figure}[ht]\par\includegraphics[scale=0.22]{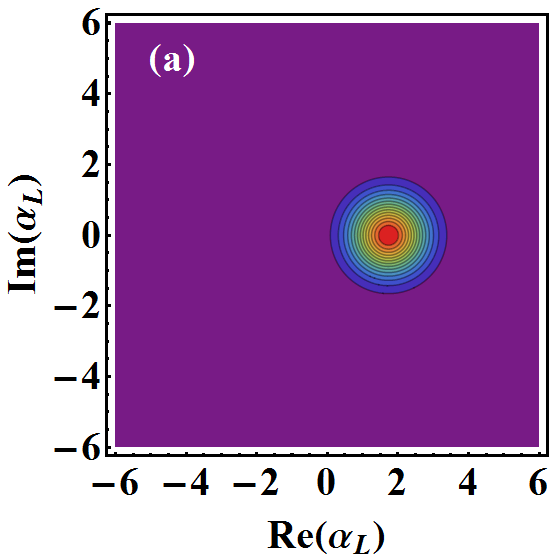}\includegraphics[scale=0.22]{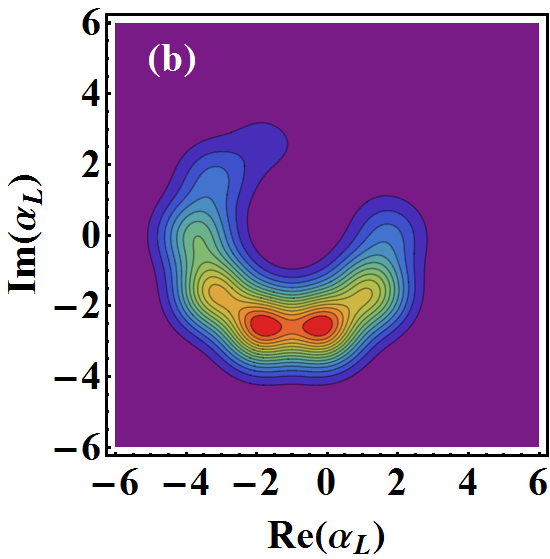}\includegraphics[scale=0.22]{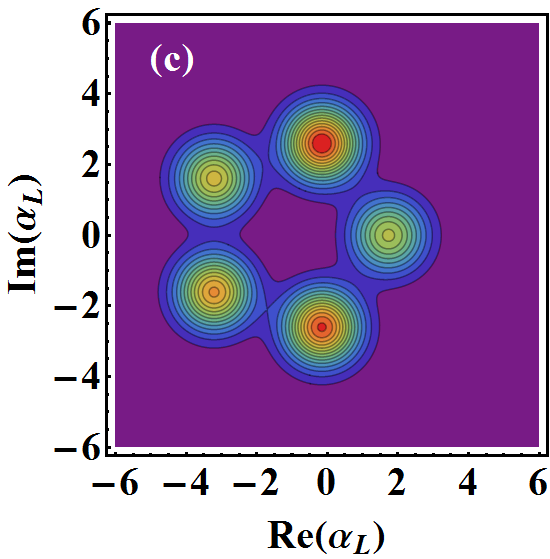}\par\includegraphics[scale=0.22]{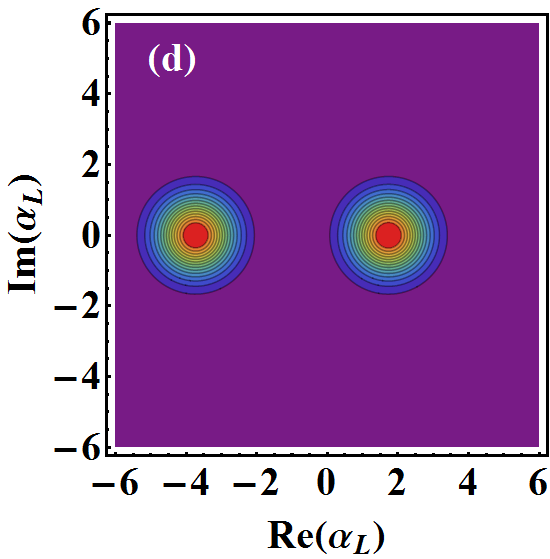}\includegraphics[scale=0.22]{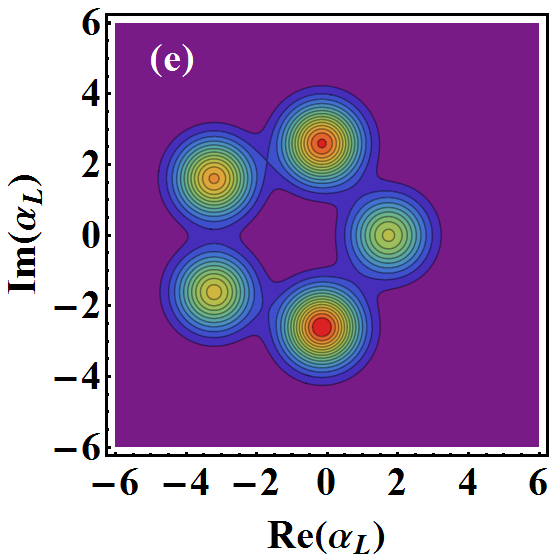}\includegraphics[scale=0.22]{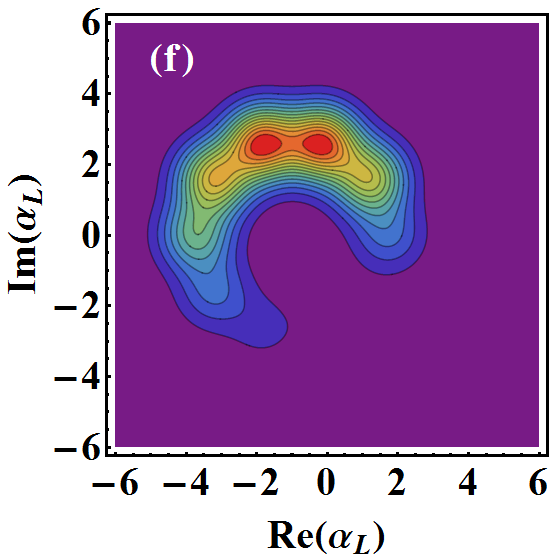}
	\caption{{Husimi function dynamics of the optical field density operator for $\vert\beta\vert^{2}=3$. We set the BEC initial state to be a coherent state whose intensity is $\vert\alpha\vert^{2}=3$. In \textbf{(a)}  $t=0$; \textbf{(b)} $t=0.1(2\pi\Delta/{\vert g_{1}\vert^{2}})$; \textbf{(c)} $t=0.4(2\pi\Delta/{\vert g_{1}\vert^{2}})$; \textbf{(d)} $t=0.5(2\pi\Delta/{\vert g_{1}\vert^{2}})$; \textbf{(e)} $t=0.6(2\pi\Delta/{\vert g_{1}\vert^{2}})$; \textbf{(f)} $t=0.9(2\pi\Delta/{\vert g_{1}\vert^{2}})$. A full revival of the initial light field coherent state in \textbf{(a)} is reached at $t^{L}_{rev}= 2\pi\Delta/{\vert g_{1}\vert^{2}}$.}}\label{Image_2}
\end{figure}

{It is interesting to note that the optical probe field will also show a collapse and revival dynamics with the same time scale, as we can see from the optical field Husimi function $Q_{L}\left(\alpha_{L},t\right)=\frac{1}{\pi}\langle\alpha_{L}\vert {\rho}_{L}\left(t\right)\vert\alpha_{L}\rangle$ depicted in Fig. \ref{Image_2}. The only extra effect that appears in the light field dynamics is due to the third term in Hamiltonian (\ref{hamiltonian}), which displaces an initial light field coherent state depending on the number of atoms in the BEC.}

{Now, by assuming a BEC initial state with a well defined number state, $\rho_{A}\left(0\right) = \vert N_{A}\rangle\langle N_{A}\vert$ (meaning $C_{m} = \delta_{m,N_{A}}$ in Eq. (\ref{evaluated_state})), the combined density operator is easily obtained and is given by $\rho\left(t\right) = \vert N_{A}\rangle\langle N_{A}\vert\otimes\vert\beta_{N_{A}}\left(t\right)\rangle\langle\beta_{N_{A}}\left(t\right)\vert$. In this case, neither the atomic nor the optical field undergo collapse and revival of their phases. From the beginning both fields are disentangled and they remain so as the initial atomic state is a eigenstate of the Hamiltonian (\ref{hamiltonian}).}
\begin{figure}[!tbp]
  \begin{center}\begin{subfigure}[b]{0.31\textwidth}
   \includegraphics[width=\textwidth]{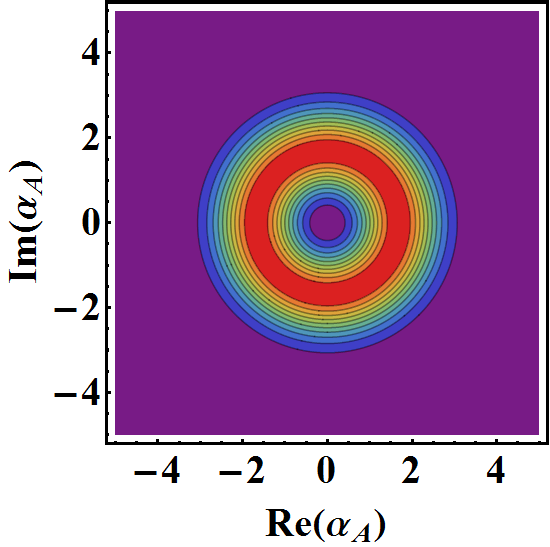}
  \end{subfigure}\end{center}
  \hspace{.46cm}\begin{subfigure}[b]{\textwidth}
    \par\includegraphics[scale=0.22]{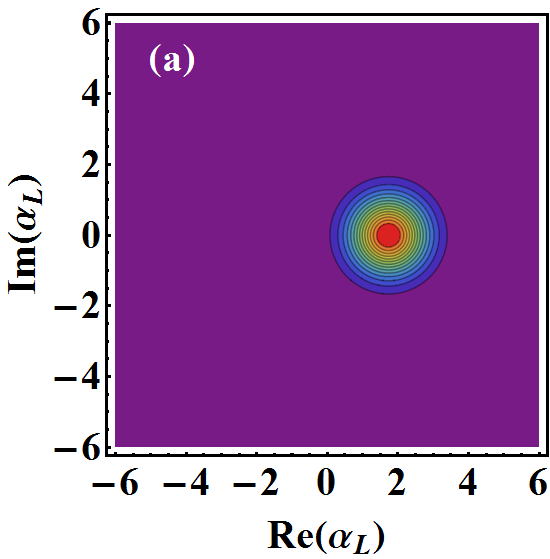}\includegraphics[scale=0.22]{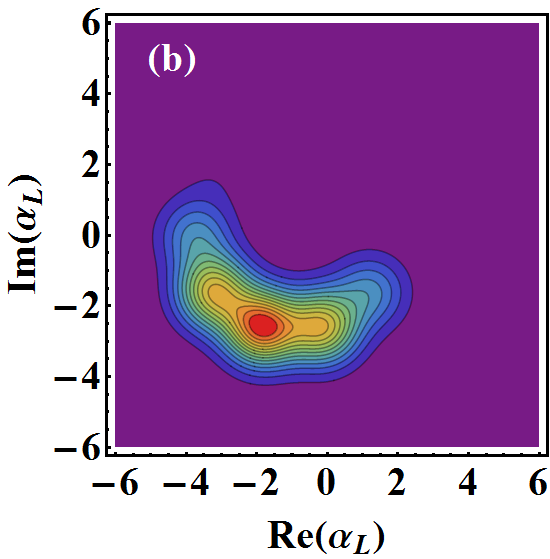}\includegraphics[scale=0.22]{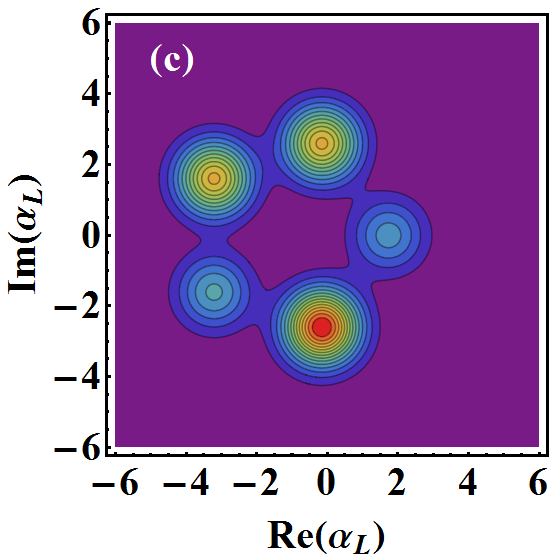}
    \par\includegraphics[scale=0.22]{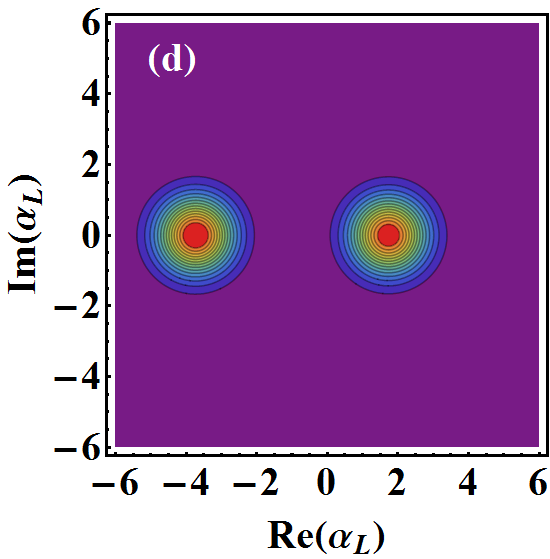}\includegraphics[scale=0.22]{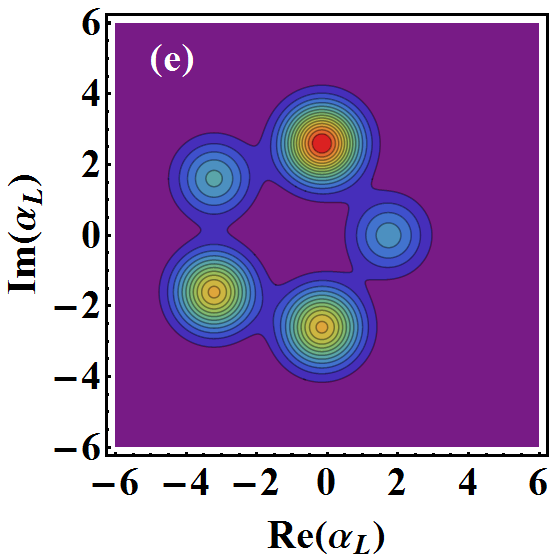}\includegraphics[scale=0.22]{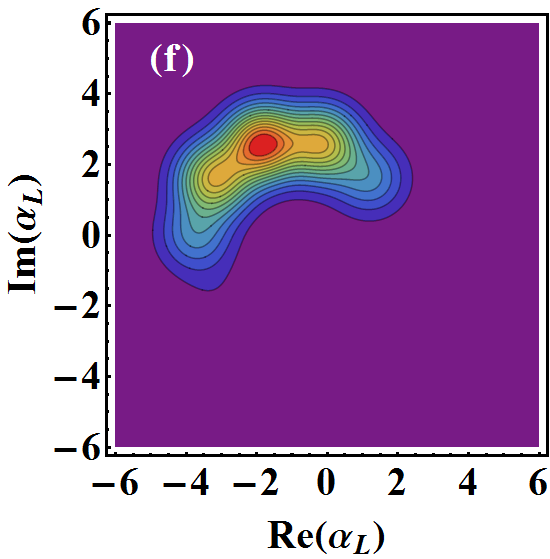}
  \end{subfigure}
  \caption{On top the atomic Husimi function for a mixed state with a binomial weight and a mean number of atoms $\langle n\rangle=3$. At the bottom, the optical field Husimi function dynamics shows collapse and revival as depicted for various time steps: in \textbf{(a)}  $t=0$; \textbf{(b)} $t=0.1(2\pi\Delta/{\vert g_{1}\vert^{2}})$; \textbf{(c)} $t=0.4(2\pi\Delta/{\vert g_{1}\vert^{2}})$; \textbf{(d)} $t=0.5(2\pi\Delta/{\vert g_{1}\vert^{2}})$; \textbf{(e)} $t=0.6(2\pi\Delta/{\vert g_{1}\vert^{2}})$; \textbf{(f)} $t=0.9(2\pi\Delta/{\vert g_{1}\vert^{2}})$; and a full revival of the initial coherent state is reached in \textbf{(a)} for $t=2\pi\Delta/{\vert g_{1}\vert^{2}}$. We set the probe field initial intensity $\vert\beta\vert^{2}=3$.} \label{Mixed_state}
\end{figure}

{
In addition we consider a statistical mixture, \begin{equation}\rho_{A}\left(0\right) = \sum_{m}P_{m}\vert m \rangle\langle m\vert.\label{mix}\end{equation} In this case the evolution of the combined system is given by
\begin{equation}
    \rho\left(t\right) = \sum_{m} P_{m}\vert m\rangle\langle m\vert\otimes\vert\beta_{m}\left(t\right)\rangle\langle\beta_{m}\left(t\right)\vert.
\end{equation}
We see that while the atoms remains in its initial mixed state, the optical field evolves as a statistical mixture of coherent states weighted by the atoms probability distribution $P_m$. This indicates that the optical field carries information about the condensate initial state. Specifically, to simplify, we consider the weights $P_{m}$ to be binomial weights
\begin{equation}
    P_{m} =\left(\frac{1}{2}\right)^{N}\left(\begin{array}{c}
     N\\
     m
\end{array}\right).
\end{equation}
State (\ref{mix}) is obtained by tracing out one mode from a two-mode BEC binomial (or atomic coherent) state \cite{John,dalton3},
\begin{equation}
    \vert\theta,\varphi\rangle = \sum_{m=0}^{N}\sqrt{\left(\begin{array}{c}
     N\\
     m
\end{array}\right)}\cos^{m}\left(\frac{\theta}{2}\right)\sin^{N-m}\left(\frac{\theta}{2}\right)e^{i\left(N-m\right)\varphi}\vert m\rangle_{1}\otimes\vert N-m\rangle_{2}, \label{atomic_coherent_state}
\end{equation}
and choosing $\theta = \pi/2$. Typically the atomic coherent state represented by (\ref{atomic_coherent_state}) is the result of the tunnelling dynamics of a BEC trapped in a symmetrical double well potential. In this situation one of the two modes works as a reference for the definition of the other mode relative phase. In Fig. (\ref{Mixed_state}) we show, through the Husimi function, the BEC and the optical field dynamics for $N=2\langle c_0^\dagger c_0\rangle =6$ (we chose $\langle c_0^\dagger c_0\rangle=N_A=|\alpha|^2=3$ to compare with the results for an initial coherent state). One can see that the optical field shows the collapse and revival dynamics, which resembles very much what is observed in Fig. \ref{Image_2}, when the BEC was prepared in a coherent state. However, since the initial BEC state now does not have a well defined phase it does not show any collapse and revival dynamics at all. Indeed it is interesting to look the comparison for all initial BEC states in  Table 1. We see that the direct inference of the BEC phase collapse and revival dynamics through the optical field behaviour is compromised, since there are situations where it shows a collapse and revival dynamics while the atomic system does not. However, although the Husimi distribution profile show a similar behaviour in both Figs. \ref{Image_2} and \ref{Mixed_state}, their structure is very different as the optical field carries information about the atomic state statistics. Therefore the observation of the optical field collapse and revival together with an analysis of its statistics allow a more complete information about the BEC state and whether it satisfies or violates the SSR. In the next section we consider a realistic continuous detection process acting on optical mode and its effects over the atomic state.}
\begin{table}
\begin{center}
\begin{tabular}{ | c | c| c | } 
\hline
\hline
\textbf{Initial atomic  state}& \textbf{BEC} & \textbf{Light}\\ 
\hline
$|N_A\rangle$ & no & no \\ 
\hline
$\sum_{m=0}^{N}P_m|m\rangle\langle m|$ & no &  yes \\ \hline
$|\alpha\rangle$& yes & yes \\ 
\hline
\end{tabular}
\caption{Collapse and revival of phase for the BEC and optical field, depending on the initial atomic state. The optical field is prepared in a coherent state $|\beta\rangle$.}
\end{center}
\end{table}
\section{Probe field photodetection}
So far we have assumed a coherent unitary evolution. Now, we turn to the situation where the probe light field is being detected. In several situations detection allows an additional control over systems \cite{saba, Diehl}. 
We employ a continuous photodetection model \cite{nosso,milburnbook} characterised by a set of operations $N_{t}\left(k\right)$, such that,
\begin{equation}
	{\rho}_{k}\left(t\right) = \frac{N_{t}\left(k\right) {\rho}\left(0\right)}{P\left(k,t\right)},
\end{equation} 
where $P\left(k,t\right) = \tr\left[N_{t}\left(k\right) {\rho}\left(0\right)\right]$ is the probability that $k$ photocounts are observed during the time interval $t$. The operation
\begin{eqnarray}
	N_{t}\left(k\right){\rho}&=&\int^{t}_{0}dt_{k}\int^{t_{k}}_{0}dt_{k-1}\int^{t_{k-1}}_{0}dt_{k-2}\dots \nonumber \\
	&&\dots\int^{t_{1}}_{0}dt_{1}S_{t-t_{k}}JS_{t_{k}-t_{k-1}}\dots JS_{t_{1}}{\rho}
\end{eqnarray}
accounts for all possible one-count process, with $J {\rho} = \gamma {a} {\rho} {a}^{\dag}$ (where $\gamma$ is the detector counting rate, which is related to the detector efficiency, including
specific physical detection process and geometrical aspects) followed by the non-unitary evolution between consecutive counts $S_{t}{\rho} = e^{Yt}{\rho}e^{Y^{\dag}t}$, with $Y = -i\frac{ {H}}{\hbar} - \frac{\gamma}{2} {a}^{\dag}{a}$. After $k$-counts on the probe field, the conditioned joint state becomes 
\begin{equation}
	{\rho}_{k}\left(t\right) = \frac{1}{P\left(k,t\right)k!}\sum_{m,n}{\rho_{A_{n,m}}\left[\mathcal{F}_{m,n}(t)\right]^{k}e^{\Phi_{m}\left(t\right)+\Phi^{*}_{n}\left(t\right)}}\left\vert m\right\rangle\left\langle n\right\vert \otimes\left\vert \beta_{m}\left(t\right)\right\rangle\left\langle \beta_{n}\left(t\right)\right\vert,
	\label{rok}
\end{equation}
 {where $\rho_{A_{n,m}}$ the matrix element of the initial state $\rho_A$ in the Fock basis, $\rho_{A}(0)=\sum_{m,n}\rho_{A_{n,m}}\vert m\rangle\langle n\vert$, and

\begin{eqnarray}
	\mathcal{F}_{m,n}\left(t\right)&=&\gamma\left\{-\frac{\Lambda_{m}\Lambda^{*}_{n}}{\Gamma_{m} + \Gamma^{*}_{n}}\left[e^{-\left(\Gamma_{m} + \Gamma^{*}_{n}\right)t} - 1\right] + G_{m}G^{*}_{n}t\right\} \nonumber\\
	&&+i\gamma\left\{\frac{G_{m}\Lambda^{*}_{n}}{\Gamma^{*}_{n}}f_n^*(t) - \frac{G^{*}_{n}\Lambda_{m}}{\Gamma_{m}}f_m(t)\right\} , \label{foda}
\end{eqnarray}
where $f_m(t)\equiv \left(e^{-\Gamma_{m}t} - 1\right)$ and 
\begin{eqnarray}
	\label{foda2}
	\Phi_{m}\left(t\right) &=& - \frac{1}{2}\left(\left|\beta\right|^{2} - \left|\beta_{m}\left(t\right)\right|^{2}\right)
	+ iG_{m}\Lambda_{m}f_m(t) \nonumber \\
	&-&\left|G_{m}\right|^{2}\Gamma^{*}_{m}t - i[(\widetilde{\omega}_{0} + \frac{\left|g_{2}\right|^{2}}{\Delta})m + \kappa m\left(m-1\right)]t. \nonumber\\
	&&
\end{eqnarray}
In Eqs. (\ref{foda}) and (\ref{foda2}), $G_{m} = \frac{g^{*}_{1}\widetilde{g}_{2}}{\Delta\ \Gamma_{m}}m$, $\Lambda_{m} = \beta + iG_{m}$, $\Gamma_{m} = i\frac{\left|g_{1}\right|^{2}}{\Delta}m + \frac{\gamma}{2}$, and $\beta_{m}\left(t\right) = \Lambda_{m}e^{-\Gamma_{m}t} - iG_{m}$. Besides $P\left(k,t\right)$ is given by
\begin{equation}
	P\left(k,t\right) = \frac{1}{k!}\sum_{m}\rho_{A_{m,m}}\left[\mathcal{F}_{m,m}\left(t\right)\right]^{k}e^{-\mathcal{F}_{m,m}\left(t\right)}.\label{probcount}
\end{equation}}
\begin{figure}[!ht]\begin{center}\includegraphics[scale=0.52]{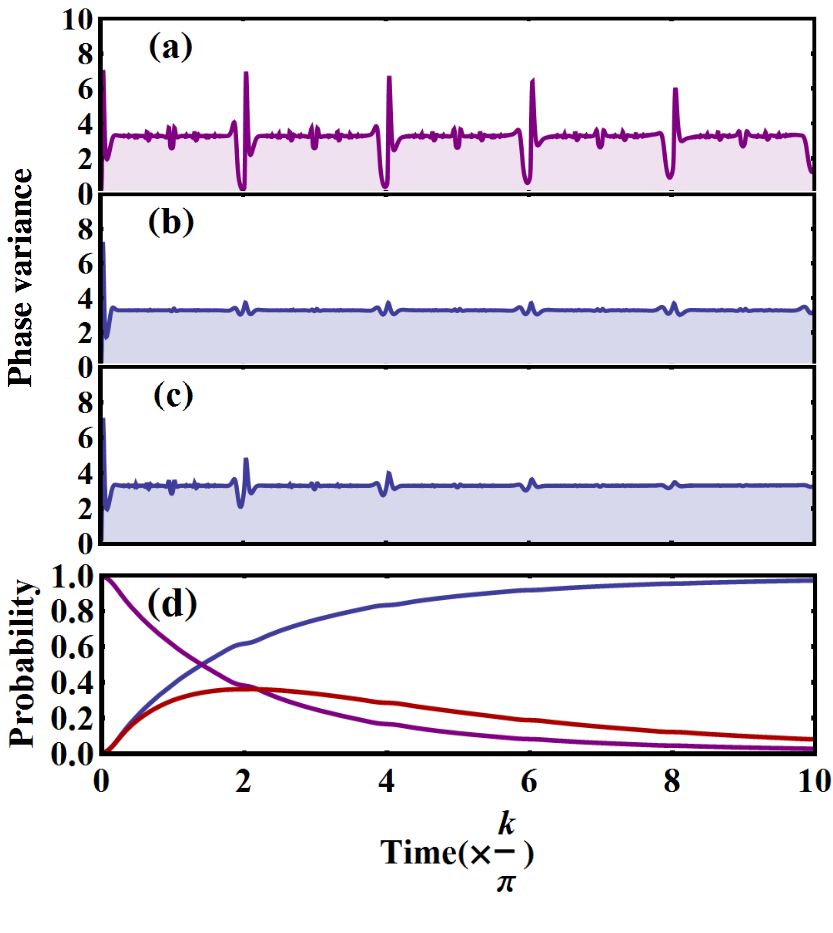}\end{center}\caption{The revival of the phase is severely affected by the counting process as depicted by the phase variance for the BEC with the photocounting process at a $\gamma = 2\times 10^{-2} \kappa$ counting rate. Other parameters are similar to the ones in Fig. \ref{phase}(f). (a) Variance for a no-count event, and (b) the variance for a single count event. (c) Variance for the pre-selected state (\ref{condstate}). (d) $P(0,t)$ (purple line), $P(1,t)$ (red line), and probability of an arbitrary counting event, $[1-P(0,t)]$ (blue line).}\label{vcount}
\end{figure}

\begin{figure}[!ht]\includegraphics[scale=0.20]{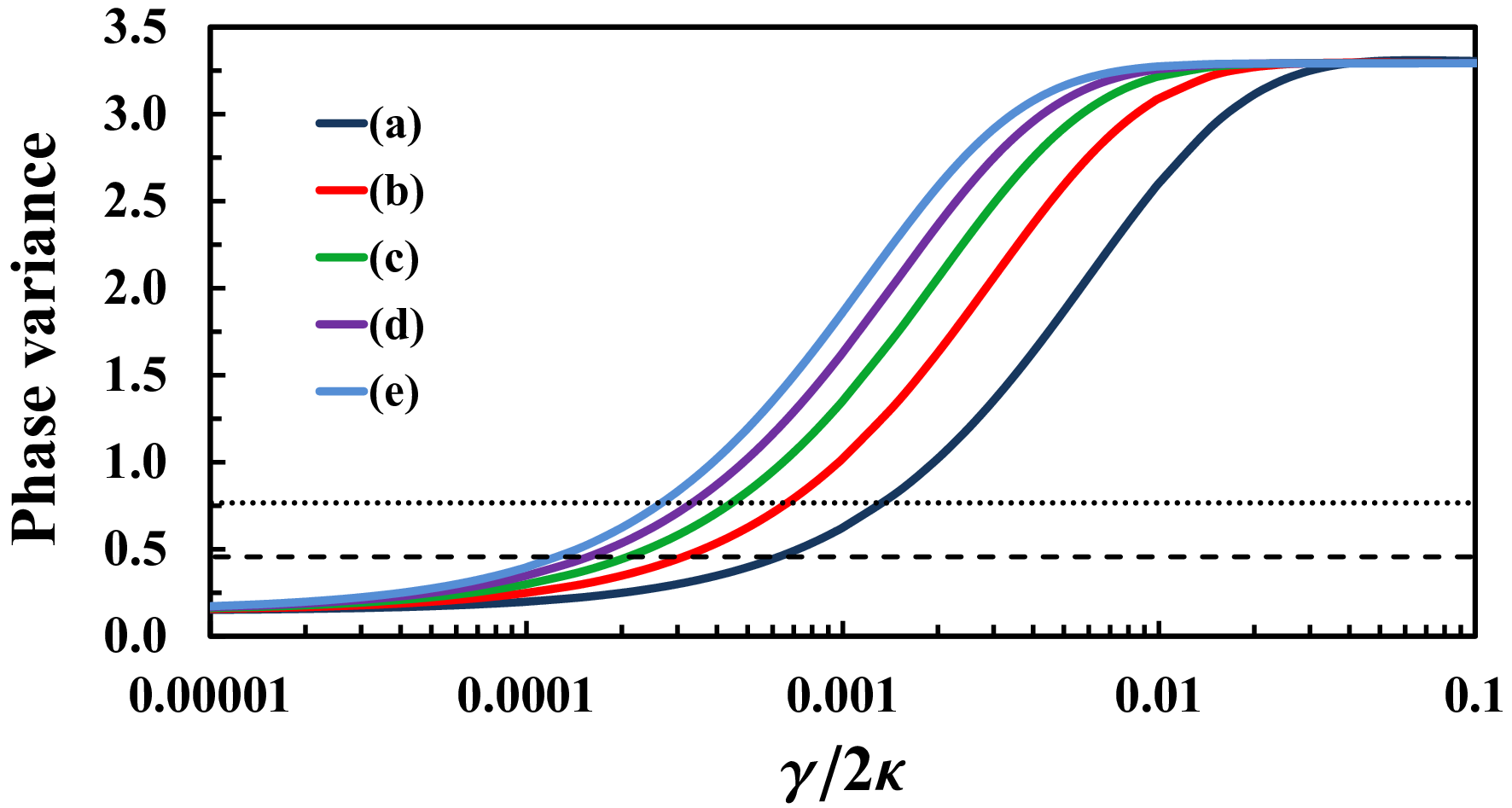}\caption{Phase variance at the revival time calculated with the pre-selected state (\ref{condstate}) as a function of the detector rate $\gamma$  for the case shown in Fig. \ref{phase}(f). \textbf{(a)}-\textbf{(e)} correspond to the first-fifth revivals, respectively. The dashed and dotted lines represent tolerance of $10\%$ and $20\%$ in the revival, respectively.}\label{vargamma}
\end{figure}

 {In Fig. \ref{vcount} we plot the phase variance and the corresponding probability of occurrence for (a) $k=0$ and (b) $k=1$, respectively, for the same parameters found in Fig. \ref{phase}(f), and with $\gamma=2\times 10^{-2} \kappa$, when the BEC is prepared in a coherent state $|\alpha\rangle$. In that situation,  $\rho_{A_{n,m}}=e^{-\vert\alpha\vert^{2}}\frac{\alpha^{m}\alpha^{*n}}{\sqrt{m!n!}}$ in Eq. (\ref{rok}), and $	P\left(k,t\right)$ is given by \begin{equation}
	P\left(k,t\right) = \frac{e^{-\frac{\vert\alpha\vert^{2}}{2}}}{k!}\sum_{m}\frac{\left(\vert\alpha\vert^{2}\right)^{m}}{m!}\left[\mathcal{F}_{m,m}\left(t\right)\right]^{k}e^{-\mathcal{F}_{m,m}\left(t\right)}.
\end{equation}

We see in Fig. \ref{vcount}(a) that the no-counting does not affect the revival of the state, at the typical time scale for a no-count event (Fig. \ref{vcount}(d)). We only see some change after a few collapse and revivals. At the revival time scale the probability that $k=1$ counting occurs is large enough so that the chance to get a revival of the BEC phase is severely compromised as we see in Fig. \ref{vcount}(b). In fact the same occurs for any $k\neq 0$. Therefore the phase evolution given by the postselected state is very sensitive, and whenever  a photodetection event occurs the whole atom-light system state is so affected that there is no chance for a revival of the initial BEC phase. This contrasts to other situations where the detection process helps to define a phase. Therefore no further control is achieved, other than the one by the detuning frequency between optical field and atoms. 
For completeness, in Fig. \ref{vcount}(c) we plot the phase variance for
unconditioned (pre-selected) state of the joint BEC-light system under the effect of counting,}
 {\begin{eqnarray}
	\hspace{-0.2cm}{\rho}(t)\hspace{-0.1cm}&=&\hspace{-0.1cm}\sum_{k}P\left(k,t\right){\rho}_{k}(t)\hspace{-0.1cm}\nonumber\\
	&=&\hspace{-0.1cm}\sum_{m,n}{\rho_{A_{n,m}}e^{\Phi_{m}\left(t\right)+\Phi^{*}_{n}\left(t\right)+\mathcal{F}_{m,n}(t)}} 
	\left| m\right\rangle\left\langle n\right|\otimes\left|\beta_{m}\left(t\right)\right\rangle \left\langle\beta_{n}\left(t\right)\right|.
	\label{condstate}
\end{eqnarray}
 This is the situation when one is absolutely ignorant about the counting events, and therefore the best one can do is to assume a convex sum of all conditioned states ${\rho}_{k}(t)$ occurring with probability $P(k,t)$. Effectively the evolved state (\ref{condstate}) is  equivalent to the situation where the light field is under the action of an amplitude damping channel, due to contact with a zero-temperature reservoir. Therefore the pre-selected state reflects the joint system incoherent evolution. 
We see in Fig \ref{vcount}(c) that there might be a chance of a partial or complete revival under damping depending on $\gamma$.
The dependence on the counting rate is better seen in Fig. \ref{vargamma} where the phase variance value at the revival times is plotted against the variation of $\gamma$ for the pre-selected state. We can see that above $\gamma/2\kappa= 6\times 10^{-4}$ ($\gamma/2\kappa= 1.2 \times 10^{-3}$) there is not a single phase revival, inside a tolerance of $10\%$ ($20\%$). All the following revivals can be tracked in a similar manner and obviously will be more sensitive to the variation of  $\gamma$.}
We remark that the observed effect for one counting event ($k=1$) would be similarly observed  had we taken an arbitrary counting event, given by the state 
\begin{equation}
	\widetilde{\rho}(t)=\frac{1}{1-P(0,t)}\left[ {\rho}(t)-P(0,t) {\rho}_{0}(t)\right],
\end{equation}
occurring with probability $1-P(0,t)$, with no further observed advantage than the one considered in Fig. \ref{vcount}. 

{However, the more strikingly use of photocounting over the probe field is the inference on the atomic BEC statistics \cite{nosso}. The $r$-th moment of the detected  $k$ photons at instant $t$ is given by
\begin{equation}
    \overline{k^r}=\sum_{k=0}^{\infty}k^r P(k,t),
\end{equation}
and with Eq. (\ref{probcount}) gives  { \begin{equation}
 \overline{k^r}=\sum_{k=0}^{\infty}\sum_{m}k^r
 \frac{1}{k!}\rho_{A_{m,m}}\left[\mathcal{F}_{m,m}\left(t\right)\right]^{k}e^{-\mathcal{F}_{m,m}\left(t\right)}.\label{av}
\end{equation}}
In the asymptotic regime $\gamma t>> 1$ for a counting rate as such $\gamma^{2}>> \frac{|g_1|^4}{\Delta^2}\langle (c_0^\dagger c_0)^2\rangle$, Eq. (\ref{av}) reduces to 
\begin{equation}
    \overline{k^r}= (\gamma t)^r \left|\frac{2g_1 \tilde{g}_2}{\gamma \Delta}\right|^{2r}\langle (c_0^\dagger c_0)^{2r}\rangle,
\end{equation}
and therefore the photocounting statistics carries information on the even moments of the condensate atom number. Let us define
\begin{equation}
\widetilde{k^r}\equiv \frac{\overline{k^r}}{(\gamma t)^r \left|\frac{2g_1 \tilde{g}_2}{\gamma \Delta}\right|^{2r}},  
\end{equation} so that the reescaled average number of counted photons is given by the second moment of atoms number in the BEC, \begin{equation}\widetilde{k}=\langle (c_0^\dagger c_0)^{2}\rangle.\label{num}
\end{equation}
Then it is immediate to check that in any of the previously considered initial state for the BEC we obtain for the average number of counted photons
\begin{equation}
\widetilde{k}_{F}=N_A^2,\;\;
\widetilde{k}_{C}=N_A(N_A+1),\;\;
\widetilde{k}_{B}=N_A(N_A+\frac{1}{2}), \label{eq37}
\end{equation}
where $F$, $C$, and $B$ stand for Fock, Coherent, and Binomial mixture, respectively and $N_A=\langle c_0^\dagger c_0\rangle$. Since the Mandel $Q$ parameter \cite{Mandel1995} for the atomic system is given by
\begin{equation}
    Q=\frac{\langle (c_0^\dagger c_0)^{2}\rangle-\langle c_0^\dagger c_0\rangle^2}{\langle c_0^\dagger c_0\rangle}-1=\frac{\widetilde{k}-N_A^2}{N_A}-1,
\end{equation}
from Eqs. (\ref{eq37}) we obtain through photodetection 
\begin{equation}
Q_{F}=-1,\;\;
Q_{C}=0,\;\;
Q_{B}=-\frac{1}{2}. 
\end{equation}
Therefore there is an enormous distinction on the statistics of the condensate state that can be probed by the optical field detection. The Mandel parameter ranges as $-1\le Q< 0$, for sub-Poissonian statistics, and $Q\ge 0$ for Poissonian and  super-Poissonian statisitcs. This means that $\widetilde{k}$ ranges as
\begin{equation}N_A^2\le\widetilde{k}< N_A\left(N_A+1\right),
\end{equation} depending on the atomic initial state being  sub-Poissonian. Also if the state is Poissonian or super-Poissonian then $\widetilde{k}\ge N_A\left(N_A+1\right)$. Consequently the analysis of collapse and revival of the optical probe field together with the photocounting statistics allows one to infer the nature of the initial BEC state. It is interesting that the nature of the initial atomic state reflects whether there is an agreement or a violation of the SSR, and so would the photocounting statistics inside the specified regimes. While it is not guaranteed that all states for whose $\widetilde{k}<N_A(N_A+1)$ satisfy the SSR, there is a strong indication that they do not if $\widetilde{k}\ge N_A\left(N_A+1\right)$, but of course this must be verified experimentally.
}

\section{Conclusion}

 The observed collapse and revival of the macroscopic matter wave field of a BEC interacting with an off-resonant quantized optical probe field is thoroughly dependent on the atom-light interaction parameters. It occurs whenever the existence of a BEC state with a well-defined phase is allowed. However for the BEC state to have a well defined phase, it relies on the existence of an observer on a reference frame from whose perspective the superselection rules  are violated. This issue of whether SSR on particle number apply or not (and in particular whether Glauber coherent states exist as physical states or are just a convenient mathematical fiction \cite{molmer}) has generated a very extensive discussion in the literature, both in the case of systems involving optical photon modes (identical massless bosons) and in the case of systems involving bosonic or fermionic atomic modes (identical massive bosons or fermions). For systems of identical massive particles there are strong reasons why it does (see \cite{leggett,dalton1,bartlett1}), whereas for systems of identical massless particles the situation is less clear (see \cite{dalton1,molmer,sanders}). However, from the point of view of a sceptic the issue is still unresolved. This issue has generated a very heated discussion in the literature, for example on whether coherent sates are only a ``convenient fiction''\cite{dalton1,dalton2,bartlett1,bartlett,molmer,sanders} or even if SSR must apply at all \cite{Susskind}, which as far as we know have no definite solution. Without entering into the merit of this discussion we offer  a more pragmatic perspective, where different sorts of initial state, in agreement or violating the SSR are assumed and  the dynamical effects over an optical probe field are thereof derived.
{When the initial BEC state has a well defined phase, a Glauber coherent state for instance, the dependence on atom-light interaction allows some degree of optical control on the atomic collapse and revival times by adjusting the coupling between the atoms and the optical field, through the variation of frequency detuning between the optical field and atoms. This collapse and revival is imprinted upon the phase of the optical probe field dynamics itself. Therefore by monitoring the dynamics of the optical probe light field one has information about the collapse and revival dynamics of the condensate. When mixtures of Fock states satisfying the SSR are assumed for the BEC, the optical phase dynamics exhibits collapse and revival effects though the BEC phase is not defined, these effects depending on atom-light field coupling constants and the atomic Fock state probabilities. When single Fock states are assumed for the BEC, no collapse and revival effects are observed in the optical dynamics. However a further analysis show that actually the optical probe carries information about the BEC state as well, which can be investigated through photocounting experiments. We show that conditioned or unconditioned continuous photodetection over the probe field does not give any additional control over the single mode BEC phase. Nonetheless it allows the  inference of the condensate state statistics through the analysis of the average number of counted photons, which together with the collapse and revival dynamics of the probe phase allows one to infer the agreement or violation of the SSR imposition on the initial BEC state. Although proposals for the preparation of single mode condensates in SSR violating states have not as yet been proposed, the system here studied can be useful for experimental verification of the occurrence of such states. Therefore, we believe that an experimental investigation along the lines of our proposal can shed some light over the issue of particle conservation SSR and its imposition on the physically allowed quantum states for many particle systems.}

\section*{Acknowledgments}{This work is partially supported by the Brazilian National Council for Scientific and Technological Development (CNPq), FAPESP through the Research Center in Optics and Photonics (CePOF) and by PNPD/CAPES.}

\appendix
\section{Derivation of Equations (\ref{evaluated_state}), (\ref{beta}) and (\ref{phi})} \label{apen}

To obtain the state given by Eq. (\ref{evaluated_state}), we apply the propagator $U\left(t\right) = \exp\left(-\frac{i}{\hbar}Ht\right)$ into the initial state given by Eq. $\vert\psi\left(0\right)\rangle = \sum_{m}C_{m}\vert m\rangle\otimes\vert\beta\rangle$ with $H$ given by (\ref{hamiltonian}). Since the first two terms in the Hamiltonian (\ref{hamiltonian}) commute with the remaining ones, we can write the propagator as follows
\begin{equation}
	U\left(t\right) = e^{-i\left[\omega_{A}n_{0} + \kappa n_{0}\left(n_{0} - 1\right)\right]t}e^{-i\left(Fa + F^{*}a^{\dag} + \xi a^{\dag}a\right)n_{0}t} , \label{appendix_1}
\end{equation}
where $\omega_{A} = \widetilde{\omega}_{0}+\frac{\left|\widetilde{g}_{2}\right|^2 }{\Delta}$, $n_{0} = c^{\dag}_{0}c_{0}$, $F = \frac{g_{1}\widetilde{g}_{2}^{\ast }}{\Delta}$ and $\xi = \frac{\left|g_{1}\right|^2}{\Delta}$. By expanding the initial atomic state in the Fock basis, the application of the propagator over the global initial state is given by
\begin{eqnarray}
	U\left(t\right)\vert\psi\left(0\right)\rangle = && \sum_{m}C_{m}e^{-i\left[\omega_{A}m + \kappa m\left(m - 1\right)\right]t}\vert m\rangle \nonumber \\
	&&\otimes e^{-i\left(Fa + F^{*}a^{\dag} + \xi a^{\dag}a\right)mt}\vert\beta\rangle . \label{appendix_2}
\end{eqnarray}
To solve $e^{-i\left[F_{m}a + F^{*}_{m}a^{\dag} + \xi_{m} a^{\dag}a\right]t}\vert\beta\rangle$, where $F_{m}= \frac{g_{1}\widetilde{g}_{2}^{\ast }}{\Delta}m$ and $\xi_{m} = \frac{\left|g_{1}\right|^2}{\Delta}m$, we employ the normal ordering method for solving Schr\"{o}dinger equation \cite{Louisel}. The generator of the evolution is the Hamiltonian of a Driven Harmonic Oscillator: $H_{DHO} = \hbar\xi_{m} a^{\dag}a + \hbar F_{m}a + \hbar F^{*}_{m}a^{\dag}$. The Schr\"{o}dinger equation
\begin{equation}
	i\hbar\frac{\partial\vert\beta\left(t\right)\rangle}{\partial t} = H_{DHO}\vert\beta\left(t\right)\rangle,
\end{equation}
has a solution given by $\vert\beta\left(t\right)\rangle = U_{DHO}\left(t,t_{0}\right)\vert\beta\left(t_{0}\right)\rangle$ where $\vert\beta\left(t_{0}\right)\rangle = \vert\beta\rangle$ and $U_{DHO}$ also satisfies
\begin{equation}
	i\hbar\frac{\partial U_{DHO}}{\partial t} = H_{DHO}U_{DHO},
\end{equation} 
subject to the initial condition $U_{DHO}\left(t_{0},t_{0}\right) = 1$. In general, a Hamiltonian $H\left(a,a^{\dag},t\right)$ in the normal order is given by $H\left(a,a^{\dag},t\right) = \sum_{l,m}h_{l,m}\left(t\right)a^{\dag l}a^{m}$, where $h_{l,m}\left(t\right)$ are \textit{c}-number expansion coefficients. The propagator for this Hamiltonian will satisfy the following equation
\begin{equation}
	i\hbar\frac{\partial U}{\partial t} = \sum_{l,m}h_{l,m}\left(t\right)a^{\dag l}a^{m}U. \label{appendix_A}
\end{equation}
Now, consider the theorem \cite{Louisel} that says if $m$ is an integer and $f\left(a, a^{\dag}\right) = f^{\left(n\right)}\left(a, a^{\dag}\right)$ (where the superscript denotes normal order), then $a^{m}f\left(a, a^{\dag}\right) = \mathcal{N}\left\{\left(\beta + \frac{\partial}{\partial\beta^{*}}\right)^{m}\bar{f}^{\left(n\right)}\left(\beta,\beta^{*}\right)\right\} = \mathcal{N}\left\{\langle\beta\vert a^{m}f\left(a,a^{\dag}\right)\vert\beta\rangle\right\}$, where $\bar{f}^{\left(n\right)}\left(\beta,\beta^{*}\right)$ is a ordinary function of the complex variable $\beta$, and $\mathcal{N}$ is an operator that transforms an ordinary function $\bar{f}^{\left(n\right)}\left(\beta,\beta^{*}\right)$ to an operator function $f^{\left(n\right)}\left(a,a^{\dag}\right)$ by replacing $\beta$ by $a$ and $\beta^{*}$ by $a^{\dag}$. With the help of this theorem we can rewrite (\ref{appendix_A}) as
\begin{equation}
	i\hbar\frac{\partial U}{\partial t} = \sum_{l,m}h_{l,m}\left(t\right)a^{\dag l}\mathcal{N}\left\{\left(\beta + \frac{\partial}{\partial \beta^{*}}\right)^{m}\bar{U}^{\left(n\right)}\left(\beta,\beta^{*},t\right)\right\} , \label{appendix_3}
\end{equation}
where $\bar{U}^{\left(n\right)}\left(\beta,\beta^{*},t\right) = \langle\beta\vert U\left(\beta,\beta^{*},t\right)\vert\beta\rangle$. If we take diagonal coherent state matrix elements of both sides of (\ref{appendix_3}) we obtain the following \textit{c}-number equation
\begin{equation}
	i\hbar\frac{\partial\bar{U}^{\left(n\right)}}{\partial t} = \sum_{l,m}h_{l,m}\left(t\right)\beta^{*l}\left(\beta + \frac{\partial}{\partial\beta^{*}}\right)^{m}\bar{U}^{\left(n\right)}, \label{appendix_4}
\end{equation}
since the right hand side is in normal order. Solving (\ref{appendix_4}), we obtain $\vert\beta\left(t\right)\rangle$ by $\vert\beta\left(t\right)\rangle = \mathcal{N}\left\{U^{\left(n\right)}\left(\beta,\beta^{*},t\right)\right\}\vert\beta\left(t_{0}\right)\rangle$. In our particular case, with the generator $H_{DHO}$, we have
\begin{eqnarray}
	i\hbar\frac{\partial\bar{U}^{\left(n\right)}}{\partial t} = && \hbar\xi_{m}\beta^{*}\left(\beta + \frac{\partial}{\partial\beta^{*}}\right)\bar{U}^{\left(n\right)} \nonumber \\
	&& + \hbar\left[F_{m}\left(\beta + \frac{\partial}{\partial\beta^{*}}\right) + F_{m}^{*}\beta^{*}\right]\bar{U}^{\left(n\right)}. \label{appendix_5}
\end{eqnarray}
If $\bar{U}^{\left(n\right)} = e^{G\left(\beta,\beta^{*},t\right)}$ where $G\left(\beta,\beta^{*},t\right) = A\left(t\right) + B\left(t\right)\beta + C\left(t\right)\beta^{*} + D\left(t\right)\beta^{*}\beta$, then (\ref{appendix_5}) becomes $i\left[\frac{dA}{dt} + \frac{dB}{dt}\beta + \frac{dC}{dt}\beta^{*} + \frac{dD}{dt}\beta^{*}\beta\right] = \xi_{m}\beta^{*}\beta + \xi_{m}\beta^{*}\left(C + D\beta\right) + F_{m}\beta + F_{m}^{*}\beta^{*} + F_{m}\left(C + D\beta\right)$, which can be separated in the following set of equations
\begin{equation}
	i\frac{dD}{dt} = \xi_{m}\left(D + 1\right),
\end{equation}
\begin{equation}
	i\frac{dB}{dt} = F_{m}\left(D + 1\right),
\end{equation}
\begin{equation}
	i\frac{dC}{dt} = \xi_{m}C + F_{m}^{*},
\end{equation}
\begin{equation}
	i\frac{dA}{dt} = F_{m}C,
\end{equation}
and whose solutions are given by $D\left(t\right) = e^{-i\xi_{m}t} - 1$, $B\left(t\right) = \frac{F_{m}}{\xi_{m}}\left(e^{-i\xi_{m}t} - 1\right)$, $C\left(t\right) = \frac{F^{*}_{m}}{\xi_{m}}\left(e^{-i\xi_{m}t} - 1\right)$ and $A\left(t\right) = \frac{\vert F_{m}\vert ^2}{\xi^{2}_{m}}\left(e^{-i\xi_{m}t} - 1\right) + i\frac{\vert F_{m}\vert ^2}{\xi_{m}}t$. Since
\begin{eqnarray}
	\vert\beta\left(t\right)\rangle &=& U\left(t\right)\vert\beta\rangle = \mathcal{N}\left\{e^{A + B\beta + C\beta^{*} + D\beta^{*}\beta}\right\}\vert\beta\rangle  \nonumber \\
	&=& e^{A\left(t\right)}e^{C\left(t\right)a^{\dag}}\mathcal{N}\left\{e^{D\left(t\right)\beta^{*}\beta}\right\}e^{B\left(t\right)\beta}\vert\beta\rangle, \label{appendix_11}
\end{eqnarray}
and $f\left(a\right)\vert\beta\rangle = f\left(\beta\right)\vert\beta\rangle$,  then
\begin{equation}
	\vert\beta\left(t\right)\rangle = e^{A\left(t\right)+B\left(t\right)\beta}e^{C\left(t\right)a^{\dag}}e^{D\left(t\right)\beta a^{\dag}}\vert\beta\rangle.
\end{equation}
By identifying $\vert\beta\rangle = e^{-\frac{\vert\beta\vert^{2}}{2}}e^{\beta a^{\dag}}\vert 0\rangle$, then we are able to recognise that
\begin{equation}
	\vert\beta\left(t\right)\rangle = e^{A\left(t\right) + B\left(t\right)\beta -\frac{\vert\beta\vert^{2}}{2}}e^{\left\{\left[1+D\left(t\right)\right]\beta +C\left(t\right)\right\}a^{\dag}}\vert0\rangle.
	\label{final}
\end{equation}
If we substitute the $A\left(t\right)$, $B\left(t\right)$, $C\left(t\right)$ and $D\left(t\right)$, $\xi_{m}$, and $F_{m}$ in (\ref{final}) we recover the Equations (\ref{evaluated_state}), (\ref{beta}) and (\ref{phi}).


\section*{References}

\bibliography{references}

\end{document}